\documentclass[aps,pra,floatfix,notitlepage,twocolumn,showpacs,preprintnumbers,longbibliography,nofootinbi,nosuperscriptaddress]{revtex4-1}

\usepackage[a4paper,left=1cm,right=1cm,top=2.5cm,bottom=2.5cm]{geometry}

\usepackage[latin1]{inputenc}
\usepackage{natbib}
\usepackage[colorlinks=true,citecolor=NavyBlue,urlcolor=NavyBlue,linkcolor=NavyBlue]{hyperref}
\usepackage[cm]{fullpage}
\usepackage{graphicx}
\usepackage[english]{babel}
\usepackage{amsmath}
\usepackage{amssymb}
\usepackage{cancel}
\usepackage{natbib}
\usepackage{comment}
\usepackage{color}
\usepackage[normalem]{ulem}
\usepackage{bbold}
\usepackage{physics} 
\usepackage[dvipsnames]{xcolor}
\usepackage{blindtext}
\usepackage{makeidx}
\usepackage{etoolbox}
\usepackage{titletoc} 

\allowdisplaybreaks

\newcommand{\stkout}[1]{\ifmmode\text{\sout{\ensuremath{#1}}}\else\sout{#1}\fi}

\begin{document}
	
\title{\textbf{\textit{Perspective}: Interactions and Nonlinearity in Non-Hermitian Physics}}
\author{Federico Roccati}
\affiliation{Quantum Theory Group, Dipartimento di Fisica e Chimica Emilio Segr\`e,
Universit\`a degli Studi di Palermo, via Archirafi 36, I-90123 Palermo, Italy}

\author{Federico Balducci}
\affiliation{Max Planck Institute for the Physics of Complex Systems, N\"othnitzer Street 38, Dresden 01187, Germany}

\begin{abstract}
    For decades, Hermiticity was considered an immutable axiom of quantum mechanics, essential for ensuring real energies and unitary evolution. This perspective has shifted radically, driven by the realization that non-Hermitian Hamiltonians provide a powerful effective description of open quantum systems, granting access to unique phenomena such as Exceptional Points and the Non-Hermitian Skin Effect. In this Perspective, we chart the trajectory of this field, moving from its established foundations in single-particle, linear models to the emerging frontier of interacting many-body systems. We first clarify the physical origins of non-Hermitian dynamics, distinguishing between mean-field approximations, conditional ``no-click'' evolution, and exact Liouvillian dynamics. We then focus on the rich phenomenology arising from the interplay of non-Hermiticity and interactions. We discuss interaction-induced topological phases, the generalization of skin effects to the many-body Hilbert space, and the distinct signatures of dissipative quantum chaos and complexity. Finally, we highlight collective phenomena in nonlinear regimes, including skin solitons and dissipative phase transitions. We also comment on measurement-induced entanglement transitions and their relation to non-Hermitian spectra and topology. By synthesizing these diverse developments, we provide a roadmap for the future of non-Hermitian physics. 
\end{abstract}

\maketitle
\tableofcontents

\section{Introduction}

This Perspective aims to chart the trajectory of non-Hermitian physics, moving from its established foundations in classical and single-particle quantum mechanics toward the complex and fertile ground of interacting many-body systems. It is not intended as an exhaustive review, but rather as a roadmap highlighting key conceptual shifts. By necessity, the cited literature is selective, focusing primarily on aspects of non-Hermitian physics that extend beyond linear and non-interacting models. For more comprehensive reviews, we refer readers to Refs.~\cite{Ashida2020, Bergholtz2021, Roccati2022}.

To provide a perspective on the field, we believe it is necessary to take a step back and look at when and in which context the interest in non-Hermitian physics sparked, highlighting some of the major milestones. Therefore, the first section is necessarily devoted to well-known results, which are amply discussed in the literature we refer to.
The second section is devoted to addressing the skepticism that often accompanies non-Hermitian physics. We clarify the physical origins of non-Hermitian effective descriptions, detailing how they arise from mean-field approximations, post-selected ``no-click'' trajectories, and the exact vectorization of the GKSL master equation. We further highlight how these theoretical frameworks have been validated by a diverse array of experiments, ranging from classical optics to superconducting quantum circuits.
The third and final section of this manuscript highlights the most recent and open frontiers of the field: those of nonlinear and interacting quantum systems. We discuss how interactions can induce topological phases in systems that lack topological properties at the single-particle level, and how the non-Hermitian skin effect manifests in the many-body Fock space. We then turn to the questions of ergodicity, reviewing recent advances in dissipative quantum chaos, many-body localization, and Krylov complexity. Finally, we explore collective phenomena, describing how nonlinearity reshapes exceptional points and drives dissipative phase transitions in many-body quantum systems.

\subsection{The Hermitian Dogma vs. Effective Reality}

The Hermiticity of the Hamiltonian stands as a cornerstone axiom of quantum mechanics, mathematically ensuring that energy spectra are real and that time evolution remains unitary, preserving the total probability~\cite{Dirac1930}. This assumption underpins the standard description of closed, isolated quantum systems, guaranteeing that dynamics are reversible.

However, this framework relies on the idealization of isolation. In experimental reality, every quantum system is inevitably open, exchanging energy or particles with an external environment. To describe this openness, one typically resorts to the rigorous theory of Open Quantum Systems, tracking the evolution of the density matrix $\rho$ via Lindblad master equations to account for decoherence and dissipation~\cite{Breuer2002}.

Alternatively, one can adopt a more compact description by retaining a Hamiltonian formalism while sacrificing Hermiticity. Historically, this was done phenomenologically by introducing complex energies, $E = E_r - i\Gamma$, to model finite lifetimes and decay~\cite{CohenTannoudji1977}. In this view, non-Hermiticity was merely a convenient mathematical approximation for leaky systems, devoid of rigorous fundamental status.

\subsection{The First Revolution: PT Symmetry}

This perspective underwent a radical paradigm shift in 1998 with the seminal work by Bender and Boettcher~\cite{Bender1998}. They challenged the necessity of the Hermitian axiom itself, posing a profound foundational question: can the Hermiticity of observables be replaced with a more physical axiom based on symmetries?
They demonstrated that a broad class of non-Hermitian Hamiltonians---specifically those respecting Parity-Time ($\mathcal{PT}$) symmetry---can exhibit entirely real spectra (a consequence of Hermiticity of observables), provided the symmetry remains unbroken~\cite{Bender1998}.

Their work marked the beginning of extensive efforts to study the mathematical and physical properties of non-Hermitian dynamics, expanding our understanding of open quantum systems. The paper transformed non-Hermitian physics from a simple tool for describing decay into a study of fundamental symmetries. It suggested that a non-Hermitian Hamiltonian could describe not just transient dissipation, but stable, balanced energy exchange. This guided the realization of physical platforms---such as coupled gain-loss waveguides---capable of probing the $\mathcal{PT}$ symmetry phase transition across Exceptional Points (EPs).

\subsection{Biorthogonal Quantum Mechanics}

From a mathematical standpoint, the new status of non-Hermitian operators necessitated a generalized framework known as biorthogonal quantum mechanics~\cite{Bender2007, Brody2013}.
For a general non-Hermitian Hamiltonian $H$, the right eigenstates $|R_n\rangle$ (satisfying $H|R_n\rangle = E_n |R_n\rangle$) and the left eigenstates $\langle L_n|$ (satisfying $\langle L_n|H = E_n \langle L_n|$) are not Hermitian conjugates of one another. Consequently, the set of right eigenstates alone is not orthogonal. Instead, orthogonality exists between the left and right sets: $\langle L_n | R_m \rangle = \delta_{nm}$ (\textit{biorthonormality}).
Completeness requires both sets, leading to the resolution of the identity $\sum_n |R_n\rangle\langle L_n| = \text{Id}$. 

A significant part of the early, more mathematically oriented literature has been devoted to ``fixing'' the non-Hermiticity of Hamiltonians possessing a real spectrum~\cite{Mostafazadeh2002}. Hermiticity is effectively a property of the inner product in the Hilbert space; if a Hamiltonian is non-Hermitian under the standard inner product $\langle \cdot | \cdot \rangle$, it may be Hermitian under a modified inner product $\langle \cdot | \cdot \rangle_\eta \equiv \langle \cdot | \eta | \cdot \rangle$. The search for such a positive-definite metric operator $\eta$ allows one to define a unitary evolution in the modified Hilbert space. For finite-dimensional systems with unbroken $\mathcal{PT}$ symmetry, such a metric can  be constructed. In the $\mathcal{PT}$-broken regime, one must resort to time-dependent metric operators to make sense of the dynamics~\cite{Fring2017}.

\subsection{How to break Hermiticity}
Having established the mathematical framework, we now turn to the physical mechanisms that generate non-Hermiticity. Consider a general Hamiltonian represented in a discrete basis as $H = \sum_{ij} H_{ij} |i\rangle\langle j|$, where the diagonal elements $H_{ii}$ correspond to onsite energies and the off-diagonal elements $H_{ij}$ represent couplings between states. Non-Hermiticity arises when the condition $H_{ij} = H_{ji}^*$ is violated, which occurs through two distinct mechanisms.

\subsubsection{Gain/Loss and Exceptional Points}
The first mechanism is the introduction of local dissipation or amplification, manifesting as complex diagonal elements (${\rm Im}(H_{ii}) \neq 0$). This can describe an open system exchanging particles or energy with a reservoir at specific sites. This mechanism is historically the playground of $\mathcal{PT}$ symmetry, where the minimal model (a gain-loss dimer) exhibits Exceptional Points (EPs)~\cite{Heiss2012}. At these spectral singularities, eigenvalues and eigenvectors coalesce, rendering the Hamiltonian defective. While EPs are often associated with the spontaneous breaking of $\mathcal{PT}$ symmetry, they are ubiquitous features of non-Hermitian parameter spaces.

\subsubsection{Non-reciprocity and Non-Hermitian Topology}
The second mechanism involves non-reciprocal couplings, $H_{ij} \neq H_{ji}^*$ $(i\neq j)$. The seminal example is the Hatano-Nelson model~\cite{Hatano1996}. In order to engineer non-reciprocity, they considered a single particle described by a Schr\"odinger equation subject to an imaginary vector potential. When this continuum model is mapped onto a discrete lattice, the imaginary gauge field modifies the hopping amplitudes resulting in asymmetric hopping rates.

Crucially,  non-reciprocal models are connected to spectral non-Hermitian topology, which is arguably the largest sub-field of non-Hermitian physics. Unlike Hermitian topology, which relies on the wavefunctions, non-Hermitian systems can possess a ``point gap'' topology where the complex spectrum winds around a reference point~\cite{Gong2018}. This winding topology is typically related to the Non-Hermitian Skin Effect (NHSE), with a recent notable exception~\cite{wang2025topologicalordernonhermitianskin}, where an extensive number of bulk modes localize at the boundaries, creating a fundamental sensitivity to boundary conditions~\cite{Yao2018}. This effect invalidates the conventional (Hermitian) bulk-boundary correspondence, and it implies that the system is extremely sensitive to the open or periodic nature of its boundaries. A broader discussion is beyond the scope of this perspective. We refer the reader to other works on the topic~\cite{Gong2018, Kawabata2019, Kunst2018,leykam2017edge} and to recent experimental implementations of dissipative non-reciprocal systems~\cite{leefmans2022topological,parto2023non,leefmans2024topological}.

\section{Non-Hermitian Dynamics and Where to Find Them}

Our discussion thus far has focused on the consistency and the unique phenomenological consequences of non-Hermitian Hamiltonians, assuming their existence as valid dynamical generators. We have discussed how breaking Hermiticity leads to a breakdown of the bulk-boundary correspondence and the emergence of spectral singularities. However, since fundamental quantum mechanics is rooted in Hermiticity to preserve probability, a crucial question arises for the skeptic: Where do these effective non-Hermitian descriptions actually originate from? Are they merely convenient mathematical artifacts, or do they describe physical phenomena? In this section, we answer this by identifying some specific physical regimes where non-Hermitian Hamiltonians arise.

\subsection{Mean-Field Dynamics}

Non-Hermitian Hamiltonians naturally emerge as effective dynamical generators in the mean-field description of open quantum systems. Consider a generic system of bosons or fermions described by a quadratic (non-interacting) Hermitian Hamiltonian $H = \sum_{ij} h_{ij} a_i^\dagger a_j$, where $a_i$ ($a_i^\dagger$) are the annihilation (creation) operators. The system is coupled to a Markovian environment via single-particle dissipators $L_{j}$, describing processes such as particle loss ($L_j = \sqrt{\gamma_j} a_j$) or gain ($L_j = \sqrt{\gamma_j} a_j^\dagger$). The evolution of the density matrix $\rho$ is governed by the standard Lindblad master equation:
\begin{equation}
    \dot{\rho} = -i[H, \rho] + \sum_j \left( L_j \rho L_j^\dagger - \frac{1}{2} \{ L_j^\dagger L_j, \rho \} \right).
\end{equation}

While the evolution of the full density matrix generally requires tracking the entire Hilbert space, for quadratic systems the dynamics of single-particle observables form a closed set of equations. Specifically, if we track the expectation values of the annihilation operators, defined as the vector $\vec{\alpha} = (\langle a_1 \rangle, \langle a_2 \rangle, \dots)^T$, a straightforward calculation shows that their time evolution is governed by a linear equation $i\partial_t \vec{\alpha} = \mathcal{H}_{\mathrm{eff}} \vec{\alpha}$.
Here, $\mathcal{H}_{\mathrm{eff}}$ is a non-Hermitian matrix containing the coherent couplings from $H$ and imaginary terms arising from the anticommutator part of the dissipator. For standard dissipation $L_j = \sqrt{\gamma_j} a_j$, the effective Hamiltonian reads $\mathcal{H}_{\mathrm{eff}} = h - i \Gamma/2$, where $\Gamma$ is a diagonal matrix of decay rates and $h=(h_{ij})$~\cite{Roccati2022, Metelmann2015}.

This derivation demonstrates that the ``non-Hermitian Hamiltonian'' is not an ad hoc ansatz but the exact equation of motion for the first moments (or coherent field amplitudes) of a quadratic open system. This formalism has been instrumental in the study of topological amplification, where the non-trivial topology of $\mathcal{H}_{\mathrm{eff}}$ ensures that edge modes are exponentially amplified~\cite{Peano2016, Porras2019, Wanjura2020, XuePRB2021}. How to physically engineer the specific non-reciprocal couplings required for these topological phases will be the subject of the next subsection.

\subsection{Engineering Non-Reciprocity}

A central challenge in realizing non-Hermitian topology is the physical implementation of asymmetric hopping. In standard passive quantum systems, hopping amplitudes are constrained by reciprocity, meaning a particle generally tunnels forward and backward with the same magnitude.

A general recipe to break  reciprocity was formalized by Metelmann and Clerk~\cite{Metelmann2015} through the concept of \textit{reservoir engineering}. The underlying principle relies on the interference between a coherent pathway and a dissipative pathway. 
%Specifically, one engineers \textit{non-local} jump operators by coupling neighboring sites to a common dissipative bath. 
Specifically, this is accomplished \textit{non-local} jump operators, 
considering a dissipator of the form $L_j = \sqrt{\kappa} (a_j + e^{i\theta} a_{j+1})$. Physically, this can be realized by coupling two primary modes (e.g., microwave cavities or optomechanical resonators) to a common, fast-decaying auxiliary mode which serves as the engineered reservoir.

When inserted into the master equation, the anti-commutator term ($- \{L_j^\dagger L_j, \rho\}/2$) generates effective non-Hermitian couplings between sites. If the system is simultaneously subject to a standard coherent hopping Hamiltonian $H = J \sum_j (a_j^\dagger a_{j+1} + \text{h.c.})$, the total effective hopping amplitudes become direction-dependent: $J_{\text{right}} = J - i e^{i\theta} \kappa /2$ and $J_{\text{left}} = J - i e^{-i\theta} \kappa/2$. Here, the phase $\theta$ controls the interference. By tuning $\theta = \pi/2$ and  $J = \kappa/2$, one achieves perfect unidirectionality ($J_{\text{left}} = 0$), thereby physically realizing the Hatano-Nelson model using standard Lindblad dynamics.

\subsection{The No-Click Limit}

The physical relevance of non-Hermitian Hamiltonians becomes even more transparent when considering the stochastic evolution of individual quantum trajectories under continuous monitoring, rather than the ensemble average~\cite{Albarelli2023}.

As an example, in the specific unraveling of continuous photodetection, the evolution consists of random quantum jumps (photon emissions) interspersed with continuous intervals of ``null results.'' Crucially, the dynamics of the system \textit{between} jumps is governed entirely by the effective non-Hermitian Hamiltonian derived earlier: $i \partial_t |\psi_c(t)\rangle = H_{\mathrm{eff}} |\psi_c(t)\rangle$, with $H_{\mathrm{eff}} = H - i\sum_j L_j^\dagger L_j/2$. In this framework, the non-Hermitian Hamiltonian is the exact generator of the conditional evolution for the subset of trajectories where no  detections are recorded. Experimentally, observing this dynamics requires post-selection: one must run the experiment many times and discard every run where the detector clicks, keeping only the ``no-click'' histories.

However, this physical interpretation faces a severe scaling challenge in the many-body limit. The probability of a trajectory surviving in the no-click manifold is given by the norm squared of the state, $P_0(t) = \| e^{-i H_{\mathrm{eff}} t} |\psi(0)\rangle \|^2$. Since the total decay rate is generally extensive, this probability vanishes exponentially with both the run time $T$ and system size $N$ ($P_0 \sim e^{-\gamma N T}$). Consequently, isolating non-Hermitian dynamics in macroscopic interacting systems becomes exponentially costly, as the required number of experimental runs diverges, often limiting this approach to small systems or short times~\cite{Ashida2020, Biella2021}. We stress that the same exponential sampling cost also plagues studies of measurement-induced phases of matter, where detecting area- and volume-law entanglement scaling requires repeated sampling from the same quantum trajectory (see also Sec.~\ref{sec:MIPT}). This issue has led some authors to argue that measurement-induced phases of matter require feedback to be operationally well-defined~\cite{Friedman2023Measurement}. Nonetheless, feedback-assisted protocols and error-mitigation/error-correction strategies may help make such trajectory-resolved signatures experimentally accessible~\cite{Kuji2026Quantum}. 

\subsection{The Vectorized Lindbladian}

While the mean-field and no-click limits rely on specific approximations or conditional unravelings, the full master equation itself provides a fundamental and unconditional source of non-Hermitian dynamics in open quantum systems.

By ``vectorizing'' the density matrix---mapping the $N \times N$ matrix $\rho$ onto a vector $|\rho\rangle\rangle$ of dimension $N^2$---the Lindblad master equation can be rewritten as a linear differential equation: $\frac{d}{dt} |\rho\rangle\rangle = \mathcal{L} |\rho\rangle\rangle$. Here, $\mathcal{L}$ is the Liouvillian superoperator. Unlike the effective Hamiltonians discussed previously, $\mathcal{L}$ includes both the coherent evolution and the full dissipator terms. Crucially, $\mathcal{L}$ is inherently non-Hermitian; its eigenvalues $\lambda_k$ are complex, with ${\rm Re}(\lambda_k) \le 0$ dictating relaxation rates and ${\rm Im}(\lambda_k)$ governing oscillatory dynamics.

It is therefore entirely legitimate to wonder whether this superoperator exhibits the unique phenomenological signatures of non-Hermitian physics. This direction has been investigated in several works. For instance, Liouvillian Exceptional Points (LEPs) have been rigorously defined and distinguished from their Hamiltonian counterparts, revealing how quantum jumps fundamentally alter the nature of spectral degeneracies~\cite{Minganti2019,minganti2020hybrid,arkhipov2020liouvillian}. Similarly, the role of symmetries in ensuring real (or paired) spectra has been extended to the superoperator level, where $\mathcal{PT}$-symmetric Liouvillian dynamics can emerge in specific operator subspaces~\cite{Prosen2012}.

Furthermore, the phenomenon of boundary localization has been generalized to the Liouvillian Skin Effect. Just as in the Hamiltonian case, the eigenmodes of $\mathcal{L}$ can pile up at the system boundaries. This has profound consequences for timescales of thermalization. It has been shown that this effect leads to ``chiral damping,'' where relaxation propagates with a sharp wavefront sensitive to boundary conditions~\cite{Song2019}. Furthermore, the Liouvillian skin effect can cause a significant slowing down of relaxation processes scaling with system size, a behavior that persists even without the closing of the spectral gap, thereby challenging standard intuitions connecting gaps to decay rates~\cite{Haga2021}. Thus, non-Hermitian topology is not merely a feature of effective Hamiltonians, but a phenomenon that manifests also in the full relaxation dynamics of open quantum systems.

\subsection{Experiments probing non-Hermitian phenomena}

The theoretical frameworks discussed above---from mean-field limits to post-selected trajectories---have moved beyond abstract speculation and found concrete realization in a diverse array of experimental platforms.

Pioneering experiments probing non-Hermitian phenomena emerged in the realm of classical optics, where the paraxial approximation of the wave equation maps formally onto the Schr\"odinger equation. In this context, experiments with coupled optical waveguides provided the first direct observation of $\mathcal{PT}$ symmetry~\cite{Ruter2010}. Soon after, numerous experiments, especially in photonics, have probed $\mathcal{PT}$ and its breaking at EPs~\cite{ozdemir2019parity,peng2014parity,peng2014loss, jing2014pt, monifi2016optomechanically,zhang2018phonon,jing2015optomechanically}. By realizing gain and loss via doping and geometric engineering, these studies observed the spontaneous breaking of $\mathcal{PT}$ symmetry and the transition from oscillatory to exponentially growing dynamics.

More recently, the focus has shifted toward probing these phenomena in fully quantum settings. Using superconducting circuits, quantum state tomography has been successfully performed across the exceptional point in a single dissipative qubit~\cite{Naghiloo2019}. Keeping only trajectories without decay events, this experiment revealed the distinct topological signature of the EP and the alteration of decoherence rates. This platform has further enabled the exploration of dynamical topology, demonstrating that encircling an EP imparts geometric phases distinct from their Hermitian counterparts~\cite{Abbasi2022}.

In the context of lattice systems, experiments have probed the unique band topology of non-Hermitian models. ``Synthetic dimensions'' in photonic ring resonators, where frequency modes are coupled via modulation, have been utilized to synthesize arbitrary non-Hermitian Hamiltonians, allowing for the direct measurement of the spectral winding number associated with ``point gap'' topology~\cite{Wang2021}. Furthermore, the macroscopic manifestations of this topology, specifically the Non-Hermitian Skin Effect (NHSE), have been observed in acoustic crystals, where sound waves in a non-reciprocal active metamaterial were shown to localize at the boundaries~\cite{Zhang2021}.

Finally, a distinct but conceptually related class of experiments involves the ``Bosonic Kitaev Chain''~\cite{McDonald2018}. In these systems, the underlying Hamiltonian is strictly Hermitian, involving both particle hopping and pair creation (squeezing). However, as discussed in Section II.A, the dynamics of the bosonic quadratures are governed by a non-Hermitian dynamical matrix. Recent realizations in multimode superconducting circuits~\cite{Busnaina2024} and nano-optomechanical networks~\cite{Slim2024} have confirmed that these systems exhibit quintessential non-Hermitian phenomena, including quadrature-dependent chiral amplification and the non-Hermitian skin effect. These results demonstrate that non-Hermitian topology dictates the physical response of quadratic bosonic systems, bridging the gap between Hermitian Hamiltonians and non-Hermitian dynamics.

\section{Interacting and Nonlinear Frontiers}

It is worth noting that much of the foundational effort in this field---from the early days of $\mathcal{PT}$ symmetry to the classification of non-Hermitian topology---was devoted to classical linear dynamics and non-interacting quantum systems. However, nonlinear behavior is ubiquitous in many physical systems, and quantum particles interact. These regimes, long considered difficult to treat, now represent the most active and fertile frontier of non-Hermitian physics.

In this section, we step into this ``interacting frontier''. We move beyond the spectral properties of simple matrices to explore how non-Hermiticity modifies, or is modified by, interparticle interactions. This regime is not merely a quantitative extension of linear models; it hosts qualitatively new phenomena. The topics we cover here are diverse and, at first glance, distinct: they range from the redefinition of topology in many-body systems to the emergence of chaos and complexity in open quantum dynamics. Yet, they share a common thread: they all probe how the interplay of non-Hermiticity and interactions/nonlinearity goes beyond the intuition developed for single-particle or closed systems.

\subsection{Topology and skin effects}
In this section, we examine how the framework of non-Hermitian topology expands when nonlinearity and many-body interactions are taken into account. While single-particle topology is well-understood, recent advances have shown that the interplay between interactions and non-Hermiticity is far from trivial. We will highlight recent advances that show how interactions can generate emergent topological phases with no linear counterpart, how topology is redefined in nonlinear driven-dissipative systems, and how the non-Hermitian skin effect manifests in the many-body regime.

\subsubsection{Interaction-Induced Topology}
A striking consequence of the interplay between non-Hermiticity and interactions is the emergence of topological phases that have no counterpart in the single-particle regime. This phenomenon, dubbed ``interaction-induced topology'' suggests that non-Hermitian topological properties can be emergent, requiring a minimum number of particles to manifest.

A pioneering realization of this concept was introduced by Faugno and Ozawa, who studied a one-dimensional lattice subject to a dynamical, density-dependent non-Hermitian gauge field~\cite{Faugno2022}. At the single-particle level, the system remains topologically trivial. However, when two particles are introduced, the interaction modifies the effective tunneling amplitudes, creating a composite ``doublon'' excitation. This composite object feels a non-reciprocal hopping that differs from its constituents, leading to the opening of a spectral point gap and the emergence of the Non-Hermitian Skin Effect exclusively in the two-particle sector.

This mechanism is not limited to explicitly non-Hermitian models but can also provide insight into Hermitian interacting systems. A recent work demonstrated that in a standard Hermitian tight-binding lattice with open boundaries, the repulsion between two particles can drive the formation of bound states whose relative motion is effectively governed by a non-Hermitian Hamiltonian~\cite{Poddubny2023}. The localization of the center of mass enforces a ``skin-like'' localization of the relative coordinate, effectively realizing an analog of the non-Hermitian skin effect induced purely by interactions. Together, these results establish that interactions can generate non-trivial non-Hermitian topology and localization phenomena that are absent in the non-interacting limit.

\subsubsection{Topology of driven dissipative nonlinear systems}

In the semiclassical regime of interacting models, the definition of topology becomes subtle. Standard band theory, which relies on the diagonalization of linear operators, cannot be directly applied to the nonlinear equations of motion governing mean-field order parameters.

A recent proposal by Villa et al.\ introduced a framework that shifts the focus from spectral topology to the topology of dynamical flows~\cite{Villa2025}. By analyzing the vector field of the semiclassical dynamics, they defined a ``graph index'' that serves as a global topological invariant. Unlike linear Chern numbers, this index characterizes the collective arrangement and nature of all fixed points (steady states) in phase space, encoding the particle-hole character of fluctuations around them.

This work bridges dynamical systems theory and topology, demonstrating that the robustness associated with topological phases translates into the structural stability of dynamical attractors. It suggests that in driven-dissipative nonlinear systems---such as networks of coupled resonators---topological transitions are intrinsically linked to bifurcations. Consequently, the classification of phases moves beyond the spectrum of a static matrix to the global geometry of the stability diagram, offering a powerful tool to predict robust operational regimes in nonlinear photonic and phononic devices.

\subsubsection{Many-body skin effects}

Beyond the single and two-particle sectors, the concept of boundary localization undergoes a fundamental transformation when applied to the full many-body Hilbert space. In this regime, the ``skin effect'' is no longer just about particles piling up at a physical edge; it manifests as an extreme localization of the system's state within the abstract geometry of the Fock space.

Recent theoretical advances have formalized this phenomenon as the Fock Space Skin Effect (FSSE). Shimomura and Sato proposed a general criterion for skin effects applicable to any linear operator, revealing that in many-body systems, the non-normality of the effective Hamiltonian can lead to eigenstates that are exponentially localized at the ``boundaries'' or ``corners'' of the configuration space~\cite{Shimomura2024}. Physically, this implies that the system becomes biased towards specific many-body configurations, leading to a drastic suppression of transport. Unlike standard thermalization, systems exhibiting FSSE display anomalously slow relaxation timescales that scale exponentially with system size, providing a robust mechanism for ergodicity breaking~\cite{Shimomura2024,HuPRL2025}.

The interplay between this localization and conservation laws yields even richer phenomenology. Gliozzi et al.\ demonstrated that in systems conserving higher-order multipoles (such as the dipole moment), the skin effect does not simply push particles to one side. Instead, it generates a ``bipolar'' localization where charges accumulate at opposite boundaries to satisfy the constraints, resulting in a state with extremal quadrupole moment~\cite{Gliozzi2024}. This structure enforces an area-law scaling of entanglement entropy even in highly excited states, challenging the standard volume-law paradigm of chaotic quantum systems~\cite{Gliozzi2024}.

Finally, interactions can drive the system towards criticality. Qin et al.\ identified a class of ``many-body critical skin effects'' arising from the competition between non-Hermitian pumping and Hubbard interactions. In this scenario, the skin effect does not affect the entire Hilbert space uniformly but emerges selectively within specific subspaces (e.g., bound states versus scattering states), driven by interaction-induced particle clustering~\cite{Qin2025}. These results collectively highlight that in the many-body interacting regime, the non-Hermitian skin effect ceases to be a mere spatial anomaly and becomes a dominant organizing principle of the quantum state in Hilbert space~\cite{MuPRB2020,KawabataPRB2022,YoshidaPRL2024}.

\subsection{Chaos, localization and complexity}

The transition to many-body systems leads also to the question of ergodicity. In closed quantum systems, the distinction between integrable and chaotic dynamics usually goes through spectral statistics: integrable systems possess Poissonian level statistics~\cite{Berry1977}, whereas chaotic systems exhibit level repulsion following the Wigner-Dyson distribution for Gaussian ensembles, as conjectured by Bohigas, Giannoni, and Schmit~\cite{Bohigas1984}. Modern diagnostics for this ``quantum chaos'' include the statistics of the level spacing ratio $r$~\cite{Oganesyan2007} and the dynamical signature known as the Spectral Form Factor~\cite{Cotler2017}.
A possible route to break ergodicity is via the phenomenon of Many-Body Localization (MBL)~\cite{Abanin2019Colloquium}: in an MBL phase, the system fails to thermalize due to the emergence of local integrals of motion~\cite{Imbrie2017Local}, and it retains memory of its initial conditions indefinitely in time.

More recently, a complementary perspective has emerged through the lens of operator growth~\cite{Barbon2019}. Here, ``Krylov complexity'' measures how a simple operator spreads over the Hilbert space under time evolution (the ``operator growth hypothesis''), serving as a refined probe of chaos and scrambling beyond standard spectral statistics~\cite{Rabinovici2025}. In the following sections, we discuss how these established paradigms have been generalized to open systems, where non-Hermiticity profoundly alters the spectral and dynamical signatures of chaos, localization, and complexity.

\subsubsection{Dissipative Quantum Chaos}

In closed quantum systems, the hallmark of chaos is the repulsion of energy levels, historically diagnosed via the statistics of real energy gaps (Wigner-Dyson statistics). In the non-Hermitian realm, eigenvalues become complex, necessitating a  generalization of these diagnostics to the complex plane.

A major step in this direction was the introduction of complex level spacing ratios by S\'a, Ribeiro, and Prosen~\cite{Sa2020}. They demonstrated that for chaotic open systems, the ratios of distances between nearest and next-nearest neighbors in the complex plane follow a universal distribution characteristic of the Ginibre ensemble. In stark contrast to the flat distribution found in integrable (Poissonian) systems, chaotic non-Hermitian spectra exhibit a distinct ``exclusion zone'' and angular anisotropy~\cite{Sa2020}. Recently, Wold et al.\ reported the experimental measurement of these statistics in a superconducting quantum processor, observing the transition from Poissonian to Ginibre distributions in a dissipative many-body system~\cite{Wold2025}.

Dynamically, quantum chaos can be captured by the Spectral Form Factor (SFF), which in closed systems displays a characteristic ``dip-ramp-plateau'' structure. The extension of the SFF to open systems reveals a competition between chaos and decoherence. Xu et al.\ showed that while dissipation generally tends to suppress or delay the formation of the ramp---the signature of spectral correlations---it does not necessarily obliterate it~\cite{Xu2021}. Interestingly, non-Hermiticity can even play a constructive role. Cornelius et al.\ demonstrated that non-Hermitian Hamiltonians with Balanced Gain and Loss can act as spectral filters. Instead of washing out chaotic signatures, this specific form of non-Hermiticity can enhance the visibility of the dip and ramp, effectively amplifying the dynamical signals of quantum chaos~\cite{Cornelius2022}. These results collectively establish that non-Hermitian physics does not merely dampen quantum chaos, but offers a new, broader playground for its manifestation.

From a broader perspective, breaking Hermiticity forces us to refine what we mean by ``quantum chaos'' in an open system. For closed systems, spectral correlations are tightly connected to long-time dynamics because unitary evolution retains information about the full spectrum: equilibration and temporal fluctuations are controlled by the distribution of many energy gaps. In contrast, non-Hermitian evolution is generically dominated at long times by the eigenmode(s) with the smallest decay rate. As a result, spectral diagnostics in the complex plane might play the same role of early-time scrambling indicators, and need not translate into a prediction on the asymptotics: a system may exhibit Ginibre-like correlations at intermediate times, yet relax to a simple stationary state. Developing a unified classification of open-system quantum chaos that reconciles spectral statistics, transient dynamics, and steady-state structure remains an active area of research.

\subsubsection{Non-Hermitian Many-Body Localization}

The introduction of dissipation challenges the stability of the Many-Body Localized (MBL) phase. The seminal work by Hamazaki and Ueda generalized MBL to non-Hermitian systems, identifying a phase transition where complex eigenvalues switch from obeying Wigner-Dyson statistics to Poissonian statistics~\cite{Hamazaki2019}. Crucially, they observed that MBL suppresses the imaginary parts of the energy spectrum, leading to a ``real-complex'' transition that stabilizes the dynamics against decay.

Following this, new diagnostic tools have been developed to better characterize this phase. Kawabata et al.\  recently introduced the use of singular value statistics, establishing their universal behavior in non-Hermitian random matrices and demonstrating their utility as a diagnostic for dissipative quantum chaos~\cite{Kawabata2023}. Building on this framework, it was shown that diagnostics based on the Singular Value Decomposition (SVD)---specifically the statistics of singular values and the entanglement entropy of singular vectors---are indicators of the localization crossover~\cite{Roccati2024,AkemannPRResearch2025}. However, a  consensus on the use of the SVD in this context has not been reached yet~\cite{BaggioliPRD2025}.

However, the ultimate fate of MBL in open systems remains a subject of intense debate. On the one hand, De Tomasi and Khaymovich provided analytical and numerical evidence supporting the existence of MBL in the ``no-click'' limit of random continuous measurements. They derived scaling laws for the critical disorder, suggesting stable localization for finite disorder strengths~\cite{DeTomasi2024}. On the other hand, recent studies focusing on dynamical transport suggest a more fragile picture. Brighi et al.\ investigated the steady-state current in non-Hermitian spin chains, finding that despite spectral signatures of localization, the system exhibits finite transport in the thermodynamic limit~\cite{Brighi2025}. Consequently, whether non-Hermitian many-body localization survives in the thermodynamic limit remains an open and actively debated question.

\subsubsection{Krylov Complexity in Non-Hermitian systems}
Recent efforts have extended the ``operator growth hypothesis''---which posits that chaotic evolution drives simple operators into complex superpositions---to the open quantum regime. A central platform for this investigation has been the dissipative Sachdev-Ye-Kitaev (SYK) model. Bhattacharjee et al.\ explored the Krylov complexity of operators evolving under a Lindbladian SYK generator~\cite{Bhattacharjee2024}. They observed that dissipation acts as a counter-force to scrambling: while the unitary part drives an exponential growth of the Lanczos coefficients (a signature of fast scrambling), the dissipative terms tend to suppress this growth. This competition defines a new dynamical regime where the saturation of complexity captures the interplay between information scrambling and environmental decoherence.

Complementing this dynamical perspective, Nandy et al.\ demonstrated the utility of Krylov subspace techniques for characterizing the statistical properties of non-Hermitian systems~\cite{Nandy2025}. By generalizing the Krylov approach to computation of the singular value decomposition (SVD), they provided an efficient method to access the singular value statistics of large-scale non-Hermitian Hamiltonians. This work reinforces the connection between the geometry of the Krylov subspace and the universal random matrix signatures of dissipative chaos discussed in the previous section.

\subsection{Nonlinearity and Collective Phenomena}

Here, we address the regime where non-Hermiticity drives macroscopic collective behavior. While the previous sections focused on statistical properties (chaos, complexity) and topological properties, here we examine how the competition between gain, loss, and interaction leads to new dynamical phases of matter. In this context, nonlinearity is not just a perturbation; it is a mechanism that can stabilize solitons against non-reciprocal drift or drive phase transitions that have no equilibrium counterparts.

\subsubsection{Nonlinear exceptional points and skin solitons}
The study of non-Hermitian singularities has traditionally focused on linear operators, where Exceptional Points (EPs) are strictly defined by the coalescence of both eigenvalues and eigenvectors (defectiveness). However, recent work has begun to unravel how nonlinearity reshapes these  concepts.

Challenging the standard definition, Bai et al.\ introduced the concept of ``nonlinear exceptional points'' (NLEPs) in dynamical systems. Unlike their linear counterparts, these singularities can exhibit a complete basis of nonlinear eigenmodes, meaning that the spectral degeneracy occurs without the  collapse of the geometric multiplicity~\cite{Bai2023}. Conversely, nonlinearity can generate exceptional topology in unexpected circumstances. Fang et al.\ demonstrated that EPs are not exclusive to non-Hermitian Hamiltonians but can emerge naturally in nonlinear Hermitian systems. In this setting, the linearization around stable equilibrium points yields an effective non-Hermitian matrix, allowing purely conservative nonlinear systems to host exceptional points and the associated phase transitions~\cite{Fang2025}.

In parallel, significant progress has been made in understanding boundary localization in the presence of interactions. In linear systems, the Non-Hermitian Skin Effect dictates that all bulk modes collapse to the edge. Nonlinearity competes with this non-reciprocal drift, leading to the formation of ``skin solitons.'' Wang et al.\ utilized a temporal photonic lattice to observe the Nonlinear Non-Hermitian Skin Effect, identifying robust soliton states that localize at boundaries due to the balance between non-reciprocal pumping and Kerr nonlinearity~\cite{Wang2025}. Extending this to higher dimensions, Kokkinakis et al.\ explored non-reciprocal lattices where nonlinearity induces a self-trapping transition. They showed that strong nonlinearity can overcome the skin effect drift, allowing for ``skin solitons'' that detach from the boundaries and stabilize in the bulk, effectively restoring bulk dynamical distinctness in systems that would otherwise be dominated by edge accumulation~\cite{Kokkinakis2025}.

\subsubsection{Phase transitions of interacting systems}

A central theme in many-body physics is the classification of phases and the transitions between them. In open systems, the interplay between unitarity, dissipation, and non-reciprocity enriches this landscape, leading to ``dissipative phase transitions'' (DPTs) that defy equilibrium classification.

Recent exact results in low-dimensional systems have shed light on the symmetry principles governing these transitions. Yao et al.\ uncovered a ``hidden time-reversal symmetry'' in boundary-driven XXZ spin chains. While the Liouvillian itself is non-Hermitian, the preservation of this hidden symmetry guarantees a steady state with specific coherence properties. The spontaneous breaking of this symmetry marks a sharp DPT, transitioning the system from a stationary phase to an oscillatory limit cycle, a behavior analytically trackable even in the strongly interacting regime~\cite{Yao2025}.

Beyond spin chains, the macroscopic coherence of bosonic condensates provides a fertile ground for non-Hermitian criticality. Belyansky et al.\ investigated driven-dissipative condensates subject to non-reciprocal hopping. They found that non-reciprocity acts as a singular perturbation to the standard superfluid: it destabilizes the uniform phase, driving a transition towards spatially modulated patterns and ``chiral'' flow states that spontaneously break translational symmetry~\cite{Belyansky2025}. Similarly, in the context of magnetism, Hanai et al.\ demonstrated that non-reciprocity can be dynamically engineered to control magnetic order. By utilizing photo-induced effective interactions, they showed that non-reciprocal exchange interactions can be generated between spins. These interactions compete with standard Heisenberg coupling, driving phase transitions between ferromagnetic and non-reciprocal magnetic phases characterized by broken inversion symmetry~\cite{Hanai2025}. These works collectively illustrate that non-Hermiticity is not merely a decay mechanism, but a control knob capable of accessing new universality classes of quantum criticality.

{
\subsection{Measurement-induced phenomena and non-Hermitian physics}
\label{sec:MIPT}
A distinctive aspect of open quantum dynamics is that dissipation can be generated by \emph{monitoring} the system: information is continuously extracted by measurements, and the ensuing back-action competes with coherent evolution. Over the last few years this viewpoint has led to the notion of \emph{measurement-induced} phases and phase transitions, most notably in the scaling of entanglement along individual quantum trajectories. Non-Hermitian physics enters this arena in a natural way: as discussed in Sec.~II.C, the conditional evolution between detection events is generated by an effective non-Hermitian Hamiltonian. This provides a direct bridge between spectral/topological structures of non-Hermitian generators and information-theoretic diagnostics such as entanglement growth and purification.

A concrete connection between measurement-induced transitions and non-Hermitian dynamics was established by Turkeshi et al.\ in the monitored quantum Ising chain~\cite{Turkeshi2021}. They compared two limits of the measurement problem: (i) a stochastic quantum-state-diffusion protocol, corresponding to a large number of weak measurement ``kicks'' per unit time, and (ii) the no-click limit, corresponding to post-selected dynamics generated by a non-Hermitian Hamiltonian. Remarkably, both protocols display a similar phenomenology upon increasing the measurement rate $\gamma$: a transition from a critical phase with logarithmic entanglement scaling to an area-law phase. The non-Hermitian Hamiltonian evolution, while it does not reproduce all the features of the (experimentally more feasible) quantum state diffusion, provides a nice theoretical explanation of the entanglement transition in terms of a spectral ``subradiance'' transition. The same transition can also explain the presence of an extended ``critical'' phase, with logarithmic scaling of the entanglement entropy, in the Ising model, which lacks a continuous symmetry. This is yet another example of how non-Hermitian Hamiltonians can lead to a richer phenomenology than their Hermitian counterparts. 

A complementary viewpoint was developed by Soares et al.\ by focusing on how symmetries and conservation laws constrain the long-time state selected by non-Hermitian evolution in free-fermionic systems~\cite{Soares2025}. They showed that, in the non-Hermitian setting, conserved quantities arise from an interplay between Hamiltonian symmetries and the initial state, due to the nonlinearity of measurement back-action. It follows that, at least in free-fermionic systems, the structure of measurement-induced steady states cannot be inferred from the spectrum alone, but also the filling of the fermionic modes must be taken into account: a ``filling-driven'' entanglement transitions between critical sub-volume scaling and area-law behavior ensues.

While measurement-induced transitions are often discussed in the context of monitored interacting circuits, non-Hermitian single-particle topology can already act as a knob to control entanglement. Kawabata et al.\ showed that the non-Hermitian skin effect can induce an entanglement phase transition in the dynamics~\cite{Kawabata2023PRX}. Indeed, the non-Hermitian skin effect generates a macroscopic particle flow, which in turn suppresses entanglement propagation, yielding area-law entanglement in the nonequilibrium steady state. The entanglement phase transition separating the area-law phase to the thermal, volume-law phase is further described by a nonunitary conformal field theory, whose effective central charge is found to be extremely sensitive to the boundary conditions. The non-Hermitian Hamiltonian formalism also allows one to pinpoint the behavior of entanglement at the transition to the presence of an exceptional point, and the corresponding structure of the eigenstates.

Taken together, these works indicate that measurement-induced phenomena provide a fertile interface between non-Hermitian spectral properties and quantum-information diagnostics. They suggest several open directions, among which extending these connections to genuinely interacting systems, and identifying experimentally viable protocols that realize these entanglement transitions without an exponentially costly post-selection overhead.
}

\section{Conclusion}

We have traced the evolution of non-Hermitian physics from a mathematical curiosity, challenging the axioms of quantum mechanics, to a rigorous framework describing open quantum systems. What began as an inquiry into $\mathcal{PT}$-symmetric Hamiltonians has matured into a comprehensive field that redefines our understanding of topology, localization, and spectral theory.

Looking forward, several key challenges and opportunities remain. First, the experimental verification of many-body non-Hermitian phenomena---such as the Fock-space skin effect or interaction-induced topology---requires a degree of control and readout fidelity that pushes the limits of current quantum simulators. Second, measurement-induced entanglement and purification transitions offer a natural meeting point between quantum trajectory physics and non-Hermitian spectra/topology, and may ultimately provide experimentally accessible probes of non-Hermitian criticality. Finally, as we master the control of dissipation, non-Hermitian dynamics may evolve from a subject of study into a tool for quantum technology, enabling new schemes for autonomous error correction, directional amplification, and robust state preparation. The ``Hermitian dogma'' has been successfully broken; the challenge now is to fully harvest the richness of the open, interacting world that lies beyond it.

\section*{Acknowledgments}

We thank Francesco Ciccarello,  Rafael Diogo Soares, Yu-Min Hu  and Franco Nori for helpful comments on the manuscript. FR acknowledges financial support by the European Union-Next Generation EU with the project ``Quantum Optics in Many-Body photonic Environments'' (QOMBE) code SOE2024\textunderscore0000084 -- CUP
B77G24000480006.

\bibliography{refs}

\end{document}